\documentclass[twocolumn]{aastex61}
\usepackage{comment}

\submitjournal{ApJ}

\newcommand{\kms}{\mbox{km s$^{-1}$}}
\newcommand{\Msun}{\mbox{M$_{\odot}$}}

\newcommand{\hii}{\mbox{ H\,{\sc ii}}}

\newcommand{\x}{\mbox{ $\times$ }}

\shorttitle{A Massive Prestellar Clump Hosting no High-Mass Cores}
\shortauthors{Sanhueza et al.}

\begin{document}


\title{A Massive Prestellar Clump Hosting no High-Mass Cores}

\correspondingauthor{Patricio Sanhueza}
\email{patricio.sanhueza@nao.ac.jp}

\author{Patricio Sanhueza}
\affil{National Astronomical Observatory of Japan, National Institutes of Natural Sciences, 2-21-1 Osawa, Mitaka, Tokyo 181-8588, Japan}

\author{James M. Jackson}
\affil{School of Mathematical and Physical Sciences, University of Newcastle, University Drive, Callaghan NSW 2308, Australia}

\author{Qizhou Zhang}
\affil{Harvard-Smithsonian Center for Astrophysics, 60 Garden Street, Cambridge, MA 02138, USA}

\author{Andr\'es E. Guzm\'an}
\affil{Departamento de Astronom\'ia, Universidad de Chile, Camino el Observatorio 1515, Las Condes, Santiago, Chile}

\author{Xing Lu}
\affil{National Astronomical Observatory of Japan, National Institutes of Natural Sciences, 2-21-1 Osawa, Mitaka, Tokyo 181-8588, Japan}

\author{Ian W. Stephens}
\affil{Harvard-Smithsonian Center for Astrophysics, 60 Garden Street, Cambridge, MA 02138, USA}

\author{Ke Wang}
\affil{European Southern Observatory (ESO) Headquarters, Karl-Schwarzschild-Str. 2, D-85748 Garching bei M\"{u}nchen, Germany}

\author{Ken'ichi Tatematsu}
\affil{National Astronomical Observatory of Japan, National Institutes of Natural Sciences, 2-21-1 Osawa, Mitaka, Tokyo 181-8588, Japan}


\begin{abstract}

The Infrared Dark Cloud (IRDC) G028.23-00.19 hosts a massive (1,500 \Msun), cold (12 K), and 
3.6-70 $\mu$m IR dark clump (MM1) that has the potential to form high-mass stars. We observed
 this prestellar clump candidate with the SMA ($\sim$3\farcs5 resolution) and JVLA ($\sim$2\farcs1 resolution) in order to characterize the early stages
 of high-mass star formation and to constrain theoretical models. Dust emission at 1.3 mm wavelength 
reveals 5 cores with masses  $\leq$15 \Msun. None of the cores currently have the mass reservoir
 to form a high-mass star in the prestellar phase.  If the MM1 clump will ultimately form high-mass stars,
  its embedded cores must gather a significant amount of additional mass over time. No molecular outflows are detected 
  in the CO (2-1) and SiO (5-4) transitions, suggesting that the SMA cores are starless. 
By using the NH$_3$ (1,1) line, the velocity dispersion of the gas is determined to be transonic or mildly supersonic
 ($\Delta V_{\rm nt}$/$\Delta V_{\rm th}$$\sim$1.1-1.8). The cores are not highly supersonic as some theories of 
 high-mass star formation predict. The embedded cores are 4 to 7 times more massive than the clump thermal Jeans 
 mass and the most massive core (SMA1) is 9 times less massive than the clump turbulent Jeans mass. 
 These values indicate that neither thermal pressure nor turbulent pressure dominates the fragmentation of MM1. 
 The low virial parameters of the cores (0.1-0.5) suggest that they are not in virial equilibrium, unless strong magnetic fields of $\sim$1-2 mG are present.
 We discuss high-mass star formation scenarios in a context based on  IRDC G028.23-00.19, a study case 
 believed to represent the initial fragmentation of molecular clouds that will form high-mass stars. 

\end{abstract}


\section{Introduction}
\label{introg}

For many decades, the study of high-mass star formation has been biased toward the more evolved, brighter, and more easily detected 
protostellar phases. In recent years, the study of the elusive prestellar phase, before the existence of embedded heating sources,
 has been recognized as key in constraining models of high-mass star formation \citep[e.g.,][]{Zhang09,Wang11,Pillai11,Tan13}. However, due to the small
  known sample of prestellar sources that have the potential to form high-mass stars ($>$8 \Msun), the current observational evidence
   is inconclusive and sometimes support or refute the same theoretical predictions. 
  
  Seen as dark silhouettes against the Galactic mid-infrared background in Galactic plane
 surveys ({\it ISO}, \citealt{Perault96}; {\it MSX}, 
 \citealt{Egan98}, \citealt{Simon06}; {\it Spitzer}, \citealt{Peretto09},
 \citealt{Kim10}), infrared dark clouds (IRDCs) are believed to host the earliest stages of star formation. 
 Several studies have investigated the kinematic and filamentary structure of IRDCs \citep{Henshaw14,Foster14,Liu14,Dirienzo15,Ragan15,Contreras16,Henshaw16}, as well as their chemistry \citep{Sanhueza12,Sanhueza13,Sakai08,Sakai12,Sakai15,Hoq13,Miettinen14,Vasyunina14,Feng16a,Kong16}.  
 Evidence of active high-mass star formation in IRDC clumps\footnote{Throughout this paper, we use 
the term ``clump'' to refer to a dense object within an IRDC with a size of 
the order $\sim$0.2--1 pc, a mass of $\sim$10$^2$--10$^3$ \Msun, and a volume density of $\sim$10$^4$--10$^5$ cm$^{-3}$. 
We use the term ``core'' to describe a compact, dense object within a clump with a size of $\sim$0.01--0.1 pc, a mass of 
 $\sim$1-10$^2$ \Msun, and a volume density $\gtrsim$10$^5$ cm$^{-3}$.} 
 is inferred by the presence of ultracompact (UC) \hii\ regions \citep{Battersby10,Avison15}, thermal ionized jets \citep{Rosero14,Rosero16},  hot cores
 \citep{Rathborne08,Sakai13}, embedded 24 $\mu$m sources \citep{Chambers09}, molecular outflows
  \citep{Sanhueza10,Wang11,Wang14,Lu15}, or maser emission \citep{Pillai06,Wang06,Chambers09,Yanagida14}. On the other hand, 
  IRDC clumps with similar masses and densities that lack all the previously mentioned star formation indicators
   are the prime candidates to be in the prestellar phase. The prestellar phase still remains the least
   characterized and understood stage of the formation of high-mass stars. 
   
   Recently, several high-mass cluster-forming clumps that are candidates to be in the prestellar phase have been 
   found mainly using {\it Herschel} observations \citep{Guzman15,Traficante15} in combination with line surveys \citep[e.g.,][]{Foster11,Jackson13,Shirley13,Rathborne16}.
    However, only few targets have been followed up in detail to 
   confirm the lack of star formation. A prestellar, high-mass cluster-forming clump completely devoid of active high-mass star formation
    \citep{Sanhueza13} and prestellar cores embedded in high-mass cluster-forming clumps \citep{Pillai11,Tan13,Beuther13,Cyganowski14,Ohashi16}
     stand out as the best candidates to study the prestellar phase in the high-mas star formation regime because they have been studied in depth using space-borne 
      telescopes, single-dish ground-base radio telescopes, and radio interferometers. Prestellar core candidates that can
       form high-mass stars have been exclusively found only toward active high-mass cluster-forming clumps.
       The prestellar core masses have a few tens of solar masses with volume densities larger than 10$^5$ cm$^{-3}$ \citep{Pillai11,Tan13,Beuther13,Cyganowski14,Ohashi16}. 
       In these cores, there is strong observational evidence 
        that turbulence alone cannot provide sufficient support against gravity to avoid rapid collapse \citep[virial parameters less than unity,][]{Pillai11,Tan13,Zhang15,Lu15,Ohashi16}. 
       Magnetic fields, which are known to play an important role in the formation of dense cores in more evolved massive clumps \citep{Zhang14}, may also be important in
        clumps at earlier stages of evolution.
       
  \subsection{High-Mass Star Formation Theories}     
 \label{intro}
       
     Current theories of high-mass star formation can be primarily separated in the way that the stars acquire their mass from the  
     environment: core accretion (``core-fed'') and competitive accretion (``clump-fed'').  The turbulent core accretion model
      \citep{McKee03} posits that all stars form by a top-down fragmentation process in which a cluster-forming clump fragments
       into cores under the combined effects of self-gravity, turbulence, and magnetic fields. These cores are gravitationally bound and 
       they are the entities that directly feed the central protostars (``core-fed''). The pressure support 
      that maintains the cores close to internal virial equilibrium is provided by turbulence and/or magnetic fields. Cores  
      have no significant further accumulation of gas from the surrounding medium, implying that the final stellar mass is smaller than 
      the core mass. The core mass is set at early times and,
       thus, in order to form a 
      high-mass star, a high-mass core must exist in the prestellar phase. Therefore, the core accretion theory predicts a direct relationship between
       the distribution function of core masses, known as the core mass function (CMF), and the mass distribution function of newly
        formed stars, known as the the initial mass function, IMF \citep{Tan14}. However, it is not clear what prevents a high-mass core 
      from fragmenting into several low-mass cores. According to \cite{Krumholz08}, the heat produced by accreting low-mass 
      stars in regions with surface densities of at least 1 g cm$^{-2}$ can halt fragmentation of the high-mass core. However, this mechanism to prevent 
      fragmentation has been questioned by \cite{Zhang09}, \cite{Smith09}, \cite{Longmore11}, \cite{Wang12}, and \cite{Tan13}. \cite{Commercon11} and \cite{Myers13} suggest that 
      magnetic fields, combined with  radiative feedback, can strongly suppress core fragmentation. However, direct observations of magnetic fields
       in very early stages of high-mass star formation remain difficult, although observations in regions with high-mass protostars appear to indicate
        that magnetic fields are important \citep{Girart09,Qiu14,Zhang14,Li15}. 
      
      Competitive accretion models posit that a cluster-forming clump fragments into cores with masses close to the thermal Jeans mass 
       \citep{Bonnell04},  $\sim$2 \Msun\ at a volume density and temperature of 5$\times$10$^4$ cm$^{-3}$ and 12 K, respectively. None of 
       these cores are massive enough to form a high-mass star. However, the cores that are located at the center of the clump's gravitational 
       potential can accrete, via modified Bondi-Hoyle accretion, sufficient mass over time to grow and eventually form high-mass stars. 
      One important distinction from the turbulent core accretion model is that the mass reservoir available to form the high-mass stars
       is accreted from material well beyond the original cores, and the gas is funneled down to the center of the clump due to the
      entire clump's gravitational potential by a large-scale infall (``clump-fed''). Thus, the core mass is gathered during the star formation process itself
        and is not set in the prestellar stage. The final stellar mass is therefore predicted to be larger than the initial core mass. According to competitive accretion, there
         are no high-mass prestellar cores, which is in
         disagreement with the turbulent core accretion model. A consequence of competitive accretion is that high-mass stars would be
          always formed near the center of stellar clusters. \cite{Krumholz05} suggest that a subvirial state is required for competitive accretion 
          to allow the formation of high-mass stars. However, \cite{Bonnell06} show that this is not the case for the simulations in \cite{Bonnell03}.  
      On the other hand, in the simulations of \cite{Wang10}, the material near the vicinity of the newly formed stars is not supported by 
      turbulence nor magnetic fields, and it is typically in a state of rapid collapse (subvirial).

     \subsection{IRDC G028.23-00.19}   
                        
           In this work, we aim to characterize the prestellar phase in the high-mass regime in order to test predictions of high-mass star 
           formation theories. \cite{Sanhueza13} studied a prestellar, high-mass cluster-forming clump candidate, MM1, located in the  IRDC G028.23-00.19 ($\sim$5,000 \Msun)
            that has the potential to form high-mass stars. The whole IRDC has been recently observed in IR polarization, which allowed the determination of magnetic field 
            strengths of 10--165 $\mu$G in the low-density regions at pc scales \citep[excluding MM1:][]{Hoq17}. 
            The massive clump  MM1 is dark at {\it Spitzer}/IRAC 3.6, 4.5, and 8.0 $\mu$m \citep{Benjamin03}, 
           {\it Spitzer}/MIPS 24 $\mu$m \citep{Carey09}, and {\it Herschel}/PACS 70 $\mu$m \citep{Molinari10}. Remarkably, MM1 has a 24 $\mu$m optical 
           depth ($\tau_{24 \mu\rm m}$) close to unity, which is the highest in the IRDC sample studied by \cite{Dirienzo15}. The total mass of the clump is 1,500
            \Msun, the radius $\sim$0.6 pc, the volume density 3\x10$^4$ cm$^{-3}$, and its distance 5.1 kpc \citep{Sanhueza12,Sanhueza13}. The clump is gravitationally
             unstable with a virial parameter significantly below unity ($\alpha$=0.3). The spectral energy distribution
             from 250 $\mu$m to 1.2 mm and the rotational diagram of low excitation CH$_3$OH lines both reveal cold dust/gas emission of 12 K \citep{Sanhueza13}. 
             Observations at 1.3, 3.6, and 6 cm that searched for free-free emission 
            \citep{Battersby10,Rosero16} and H$_2$O and CH$_3$OH maser emission \citep{Wang06,Chambers09} have resulted in null detections. The cold temperatures,
             coupled with the lack of several indicators of star formation, suggest that the massive clump MM1 is a pristine prestellar clump appropriate for the study of the earliest
              stages of high-mass star formation. By using high-angular resolution observations from SMA ($\sim$3.5\arcsec\ resolution) and JVLA ($\sim$2\arcsec\ resolution), we 
              have searched for the embedded cores and determined their dynamical state at $<$0.1 pc scales.

\section{Observations}

Observations of IRDC G028.23-00.19 were carried out with the Submillimeter
 Array\footnote{The Submillimeter Array is a joint project between the Smithsonian
Astrophysical Observatory and the Academia Sinica Institute of Astronomy
and Astrophysics, and is funded by the Smithsonian Institution and the
Academia Sinica} (SMA) and the Karl G. Jansky Very Large Array (JVLA), operated by 
the National Radio Astronomy Observatory\footnote{The National Radio Astronomy
 Observatory is a facility of the National Science Foundation operated under cooperative
  agreement by Associated Universities, Inc.}. 

\subsection{SMA Observations}
\label{sma-obs}

SMA 1.3 mm  line and continuum observations were taken during April 2012 and July 2013 in the
 compact configuration. The projected baselines range from 10 to 69 m.
The IRDC was completely mapped by combining images in a mosaic from 5
 separate positions, using the same correlator 
setup which covers 4 GHz in each of the lower and upper sidebands.
A spectral resolution of 1.1 \kms\ (0.812 MHz) was used. The continuum emission
 was produced by averaging the line-free channels in visibility space.  
Using natural weighting, the 1$\sigma$ rms noise for the continuum emission 
is 0.75 mJy beam$^{-1}$. 

The system temperature typically varied from 150 to 220 K during the
 observations. At the center frequency of 224.6975 GHz (1.3 mm), the primary
 beam or field of view of SMA is 56\arcsec. These SMA observations are 
sensitive to structure with angular scales smaller than $\sim$30\arcsec.  
The final synthesized beam has a size of 4\farcs1$\times$3\farcs0 with a P.A. of
 -25\arcdeg. The geometric mean of the major and minor axis is 3\farcs5, which
 corresponds to a physical size of $\sim$0.09 pc ($\sim$18,000 AU) at a 
 distance of 5.1 kpc \citep{Sanhueza12}. 

 The data from different tracks were calibrated separately using the
 IDL-based MIR package and exported to CASA to be combined in the visibility
 domain for imaging. Typical SMA observations may be subject to up to 15\% of
 uncertainty in absolute flux scales. The quasar J1743-038 was periodically 
observed for phase calibration. The quasars J2202+422 (BL Lac) and
 J0319+415 (3C84) were used for bandpass calibration. Uranus and the bright
 radio continuum source MWC349A were used for flux calibration. 

\subsection{JVLA Observations}

The JVLA observations consisted of two pointings at K-band (1.3 cm)
 covering the whole IRDC, taken in February 
 2012 in the C configuration. The projected baselines ranged from 50 to 3,400 m. 
NH$_3$ (J, K)=(1, 1) at 23.6944955 GHz and NH$_3$ (J, K)=(2, 2) at 23.7226336 
GHz were simultaneously observed with a spectral resolution of 0.4 \kms\ 
(31.25 kHz) in dual polarization mode. At these frequencies, the primary
 beam of JVLA is 1\farcm9. These observations are
 sensitive to angular scales smaller than $\sim$1\arcmin. 

The data were calibrated and imaged using the CASA 4.2 software package. 
In order to improve the S/N ratio, a Gaussian filter ``outertaper'' of
 1\farcs6 was applied during the {\sc clean} deconvolution, obtaining twice the
 original synthesized beam. Using natural weighting, the synthesized beam was  
2\farcs3$\times$2\farcs0 with a P.A.=32\arcdeg\ for both NH$_3$ lines, while  
the 1$\sigma$ rms noise was 0.78 mJy beam$^{-1}$ per channel for NH$_3$ (1,1)
 and 0.75 mJy beam$^{-1}$ per channel for NH$_3$ (2,2). The conversion factor
 between mJy beam$^{-1}$ to brightness temperature in K is 0.47
 (1 mJy beam$^{-1}$ = 0.47 K). The geometric mean of the major and
 minor axis is 2\farcs1, which 
 corresponds to a physical size of $\sim$0.05 pc ($\sim$11,000 AU) at a 
 distance of 5.1 kpc.

The bandpass and flux calibrations were performed by using 
observations of the quasar J1331+305 (3C286). The phase calibration was 
done by periodically observing the quasar J1851+0035.

\section{Will the clump MM1 in IRDC G028.23-00.19 form High-Mass Stars?}

Current observational evidence supports the idea that the clump MM1 will form high-mass
stars. In addition, the clump properties are also consistent with those used in 
simulations that produce stellar clusters including high-mass stars. We assess 
the potential to form high-mass stars of the IRDC G028.23-00.19 below:

i) Based on the observed relation between the maximum stellar mass in a
 cluster ($m_{\rm max}$) and the total mass of the cluster ($M_{\rm cluster}$), given by \citep{Larson03}
 \begin{equation}
\left(\frac{m_{\rm max}}{\Msun}\right) = 1.2\,\left(\frac{M_{\rm cluster}}{\Msun}\right)^{0.45}~, 
\label{larson-eq}
\end{equation}
  the mass of the most massive
 star that will be formed in a cluster-forming clump can be estimated. 
 Assuming a cluster star formation efficiency of 30\% \citep{Alves07,Lada03},
the clump MM1 in IRDC G028.23-00.19, which has a mass of 1,500 \Msun, should
 form a stellar cluster of a total mass of 450 \Msun. According to the discussion in 
 Section~\ref{sec:error}, the uncertainty in the mass is 50\%. Larson's relationship
 then predicts that this clump will form one high-mass star of
 19 $\pm$ 4 \Msun. Even with only 5\% star formation efficiency, a 8 $\pm$ 2 \Msun\  
 star should be formed.

Using the IMF from \cite{Kroupa01}, the maximum stellar mass of a clump 
can also be estimated and is given by (see Appendix~\ref{app})
\begin{equation}
m_{\rm max} = \left(\frac{0.3}{\epsilon_{\rm sfe}}\frac{17.3}{M_{\rm clump}} + 1.5\times 10^{-3}\right)^{-0.77}~.
\label{eqn-IMF-m-max}
\end{equation}

For the clump MM1, assuming 30\% star formation efficiency ($\epsilon_{\rm sfe}$), 
a high-mass star of 28 $\pm$ 9 \Msun\ can be formed.

ii) \cite{Kauffmann10} find an empirical high-mass star formation threshold,
 based on clouds with and without high-mass star formation. They suggest
 that IRDCs with masses larger than the mass limit given by $m_{\rm lim}$ =
 580 \Msun\ (r/pc)$^{1.33}$, where $r$ is the source radius, are forming high-mass stars or will likely form them 
 in the future. To be consistent with calculations made in our work, we use the factor 
 of 580 directly obtained by using the \cite{Ossenkopf94} dust opacities and not the 
 decreased (by a factor 1.5) dust opacities that lead to the original value of 870 
  \citep[see][for details]{Kauffmann10}. Applying the \cite{Kauffmann10} relationship
   to the clump MM1 (r $\approx$ 0.6 pc), its 
 corresponding mass threshold is 290 \Msun, well below the measured mass of
 1,500 \Msun. Thus, ``the compactness'' ($M_{\rm dust}$/$m_{\rm lim}$) of MM1
 is 5.2 and it is highly likely that the clump will form high-mass stars. 

Based on large samples of high-mass star-forming regions, \cite{Urquhart14} and \cite{He15} 
 suggest that high-mass stars are formed in clumps with $\Sigma_{\rm clump} >$ 0.05 gr cm$^{-2}$. 
 \cite{Lopez10} suggest a significantly larger surface density threshold 
 (0.3 gr cm$^{-2}$) for high-mass star formation based on the detection of
  massive outflows (outflow mass $>$ 10 \Msun). MM1 has $\Sigma_{\rm clump}$ and $\Sigma_{\rm peak}$ of 
  0.3 and 0.4 gr cm$^{-2}$, respectively. The surface density in MM1 was calculated as $\Sigma$ = M/($\pi r^2$) with 
  $r_{\rm clump}$ = 0.6 pc and $r_{\rm peak}$ = 0.14 pc using the IRAM 30 m dust continuum observations at 
  11\arcsec\ angular resolution (see Figure~\ref{SMA_IRAM}). The surface densities are significantly larger than the lower 
 limit suggested for high-mass star formation and consistent with the highest threshold. 

With a low virial parameter \citep[$\alpha$ = 0.3,][]{Sanhueza13}, IRDC G028.23-00.19 MM1 is unstable and 
subject to collapse. Therefore, several pieces of evidence 
 indicate that IRDC G028.23-00.19 will almost certainly form a stellar cluster including high-mass
  stars. The question that remains open is how. Since IRDC G028.23-00.19 MM1
   appears to be in the prestellar stage and is likely to form high-mass stars, observations
    of this massive clump can be used to distinguish the early stage differences posited by 
    high-mass star formation theories (see Section~\ref{intro}).

\section{Results}

\subsection{Dust Continuum Emission}

\begin{figure*}
\begin{center}
\includegraphics[angle=0,scale=0.2]{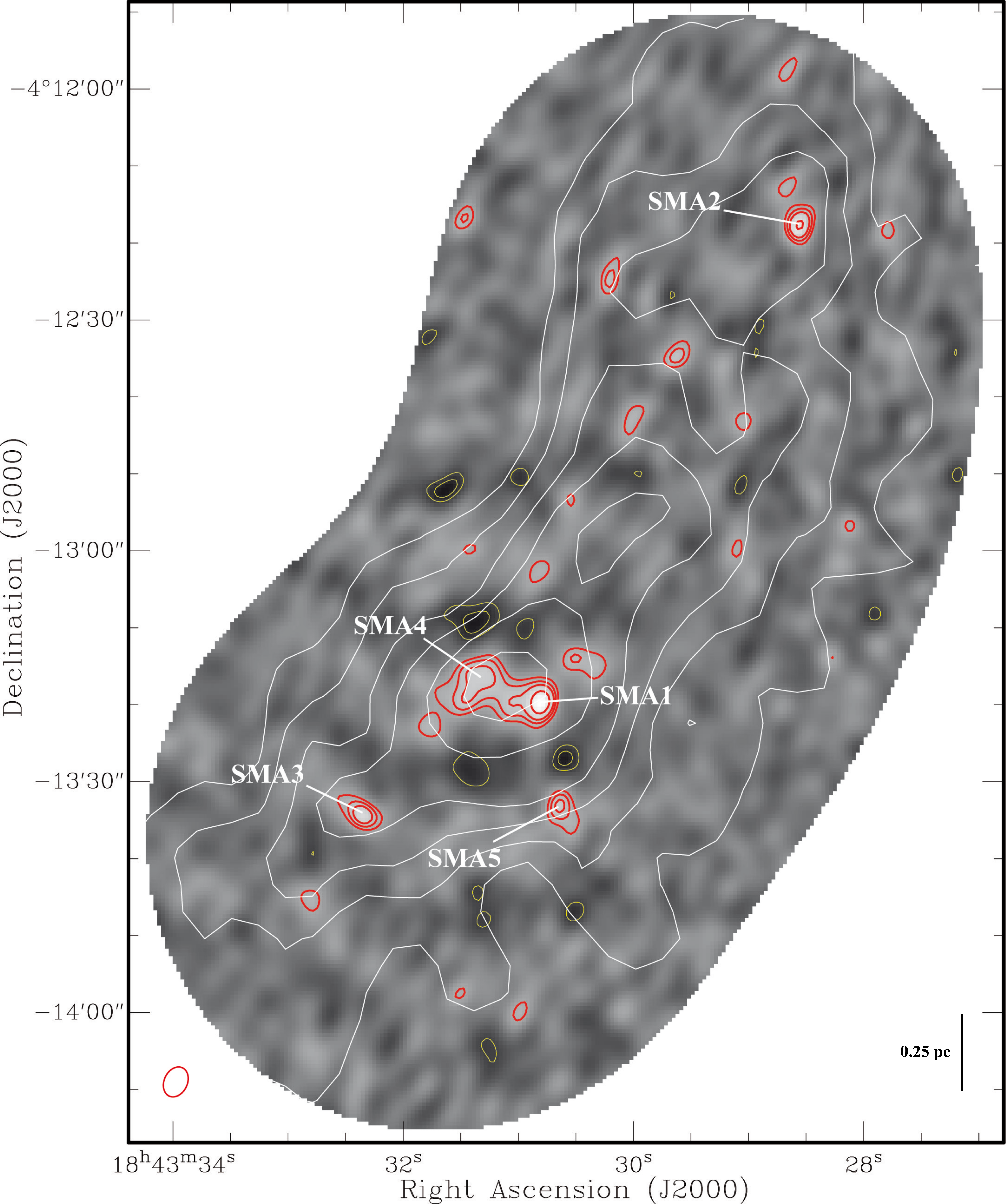}
\end{center}
\caption{SMA 1.3 continnum emission ($\sim$3\farcs5) in gray-scale image and red
 contours of 
 IRDC G028.23-00.19 overlaid with 1.2 mm continuum emission from the IRAM
 30 m telescope (11\arcsec) in white contours. Contour levels for the 
1.3 continnum emission are -4, -3, 3, 4, 5, and 7 $\times$ $\sigma$, with 
$\sigma$ equal to 0.75 mJ beam$^{-1}$. Contour levels for the 1.2 mm
 continuum emission are 20 ($\sim$3$\sigma$), 35, 50, 65, 85, and 105 mJy
 beam$^{-1}$. The position of the five cores detected with SMA above
 5$\sigma$ are shown. The synthesized SMA beam is displayed at the bottom left of 
the image.}
\label{SMA_IRAM}
\end{figure*}

Figure~\ref{SMA_IRAM} shows the 1.3 mm dust continuum emission from SMA
 ($\sim$3\farcs5 angular resolution) in gray scale and red contours.
 White countours correspond to
 the 1.2 mm dust continuum emission from the single-dish IRAM telescope
 (11\arcsec\ angular resolution). In the whole cloud, five SMA cores are
 detected above 5$\sigma$, which corresponds to 4.5 \Msun\ (following the procedure 
 described in Section~\ref{sma-mass}). 
The 5$\sigma$ threshold was used because the sidelobe pattern produces negative
 artifacts with absolute values as large as 4$\sigma$, 
suggesting that some of the positive detections at 3$\sigma$ and
 4$\sigma$ may be spurious sources due to side lobes instead of real sources. The cores are
 named SMA1, SMA2, SMA3, SMA4, and SMA5 in the order of decreasing peak
 flux (see Figure~\ref{SMA_IRAM} and Table~\ref{tbl-SMA-cores}).
Except for SMA4, all cores were fitted by 2-D Gaussians in the CASA software 
package. The fitted
 parameters are listed in Table~\ref{tbl-SMA-cores}. The deconvolved size 
was adopted  to determine the physical size of the cores. No 2-D Gaussian
 fit succeeded for SMA4. The flux inside the countour defined at the 
4$\sigma$ level was used to estimate its integrated flux. To estimate the physical
 parameters of SMA4, we adopted the synthesized beam as the physical size. The parameters for 
SMA4 should be treated with caution because this core could be composed
 of a few unresolved condensations. SMA4 is not centrally peaked as the other 
 SMA cores and it approximately has a constant brightness above the 5$\sigma$ contour. 

 From \cite{Rathborne10}, the high-mass clump MM1 has a 1.2 mm 
single-dish integrated flux of 1.63 Jy. Comparing this integrated flux with
 the 1.3 mm SMA integrated flux of $\sim$80 mJy,  $\sim$8\% of the
 single-dish flux is recovered by the interferometer (assuming
 $\beta=1.8$ to compare the somewhat different frequencies).

\begin{deluxetable*}{lcccccc}
\tabletypesize{\normalsize}
\tablecaption{SMA Core Parameters \label{tbl-SMA-cores}}
\tablewidth{0pt}
\tablehead{
\colhead{Core} &  \multicolumn{2}{c}{Position}  & \colhead{Peak Flux}
 & \colhead{Integrated Flux} & \colhead{Angular Size}& \colhead{Deconvolved Size}\\
\colhead{Name} &  \colhead{$\alpha$(J2000)} & \colhead{$\delta$(J2000)}& \colhead{(mJy beam$^{-1}$)}&\colhead{(mJy)}&\colhead{(\arcsec$\times$\arcsec)}&\colhead{(\arcsec$\times$\arcsec)}\\
}
\startdata
SMA1 & 18:43:30.81 & -04.13.19.6 & 8.18 & 12.2 & 4.8 $\times$ 3.8 & 2.8 $\times$ 2.1\\
SMA2 & 18:43:28.55 & -04.12.17.6 & 8.14 & 10.2 & 4.5 $\times$ 3.4 & 2.2 $\times$ 1.1\\
SMA3 & 18:43:32.36 & -04.13.34.3 & 5.35 & 7.04 & 4.5 $\times$ 3.6 & 2.0 $\times$ 1.9\\
SMA4 & 18:43:31.31 & -04.13.16.4 & 5.29 & 9.28  & \dots & \dots \\
SMA5 & 18:43:30.64 & -04.13.33.1 & 4.80 & 6.97 & 5.0 $\times$ 3.6 & 3.1 $\times$ 1.5 \\
\enddata
\tablecomments{Fitting uncertainties are $<$1\% for the peak flux, integrated flux, and the angular size, and $<$3\% for 
the deconvolved size. For the calculation of physical properties, as discussed in Section~\ref{sec:error}, the uncertainties of flux and size measurements are 
dominated by  the absolute flux scale of SMA (15\%) and the distance to the source (10\%). No 2-D Gaussian fit was reliable for SMA4. In order to estimate 
its total flux, the flux inside the contour defined at 4$\sigma$ was
 integrated. Its adopted size is the SMA synthesized beam. } 
\end{deluxetable*}

\subsection{SiO and CO Emission}

High velocity gas in the SiO and CO lines is frequently interpreted as molecular
 outflows, and thus they can reveal deeply embedded active star formation that
  can be undetected at IR wavelengths. We searched for emission  from the SiO (5-4) transition and 
  found no detection in IRDC G028.23-00.19 MM1 at a sensitivity of 38 mJy beam$^{-1}$ per channel of 1.1 \kms. 
  The emission from the CO (2-1) transition is detected but the line profiles are heavily affected by self-absorption and/or 
  missing flux. At a sensitivity of 40 mJy beam$^{-1}$ per channel of 1.1 \kms, there is 
  no evidence of wing emission that indicates protostellar outflows. Therefore, 
  at the sensitivity level of these observations, we confirm that the cores embedded in IRDC G028.23-00.19 
  MM1 clump are starless. \cite{Sanhueza13} found SiO (2-1) emission to the north and south of MM1. With the SMA 
  observations, we confirm the absence of molecular outflows and the more likely mechanisms for releasing SiO to the gas phase 
  are large scale shocks rather than active star formation.

\subsection{NH$_3$ Emission}
\label{nh3_emission}

\subsubsection{Images}

\begin{figure*}
\begin{center}
\includegraphics[angle=0,scale=0.8]{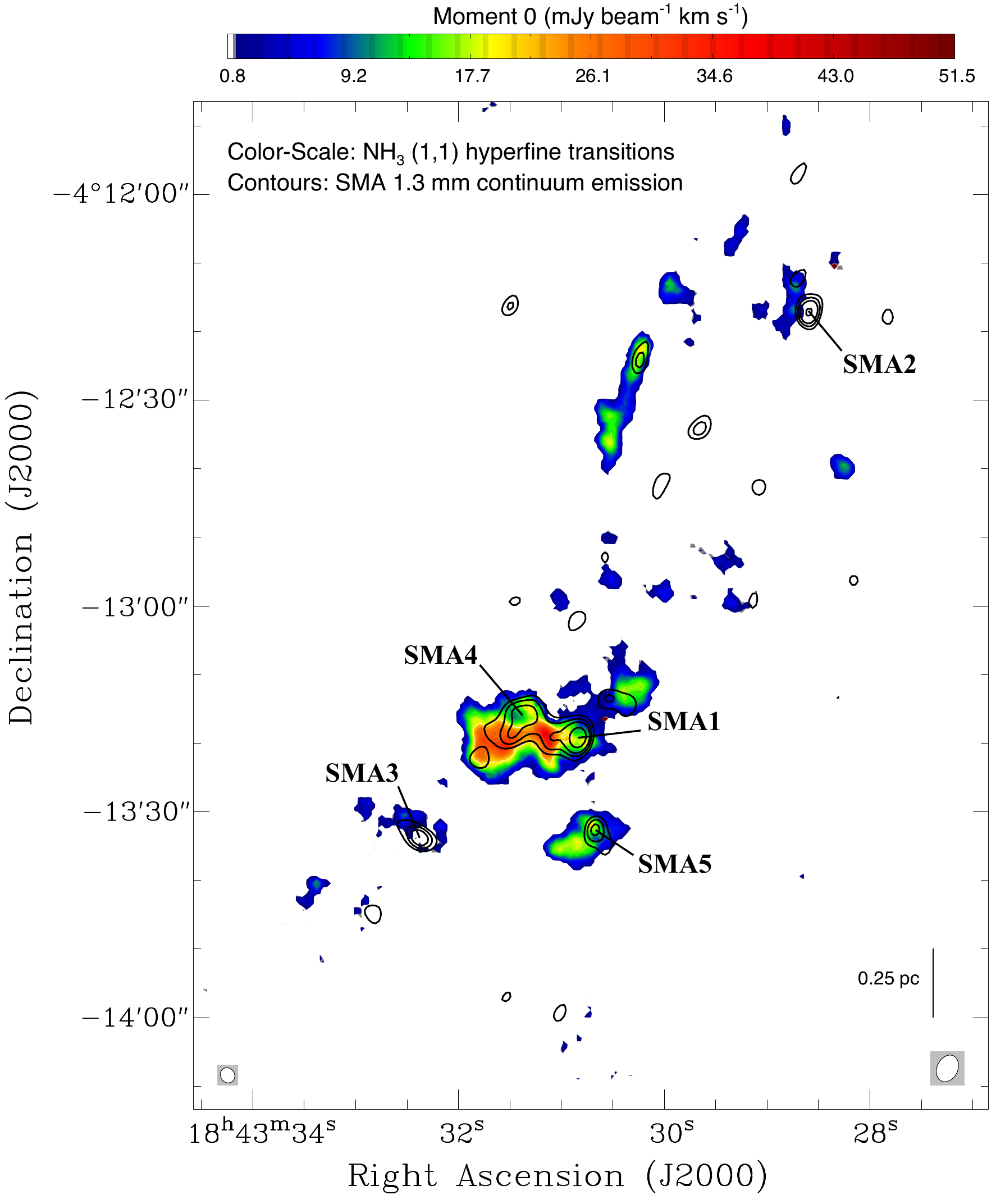}
\end{center}
\caption{NH$_3$ moment 0 map (integrated intensity) in
 color-scale  overlaid with 1.3 mm continuum emission from SMA in black
 contours. The moment 0 map includes the data above 2.5$\sigma$ per channel
 for all five hypefine lines. All emission above 15\% of the peak emission is
 shown. Contour levels for the 1.3 continnum emission are 3, 4, 5, and
 7 $\times$ $\sigma$, with $\sigma$ equals to 0.75 mJ beam$^{-1}$.
 Angular resolutions of JVLA (2\farcs1) and SMA (3\farcs5) are shown in the
 bottom left and right, respectively.}
\label{mom0_1.3mm}
\end{figure*}

Figure~\ref{mom0_1.3mm} shows, in color scale, the
 moment 0 (integrated intensity) map of the five NH$_3$ hyperfine lines
 overlaid with the 1.3 mm dust continuum emission from SMA. The 
 data above 2.5$\sigma$ per channel were used for making the moment map.
NH$_3$ emission is generally associated with dust emission, although the 
molecular emission is more spatially extended than the dust emission and 
is sometimes detected in regions without a dust counterpart.

The global, large-scale kinematics of the IRDC will be studied in detail in a following paper 
in which we will recover the missing flux by combining the NH$_3$ JVLA 
 interferometric observations with the single-dish
 Green Bank telescope (GBT) observations. In this paper, we focus on the compact cores detected
  in the prestellar, high-mass clump at the center of the IRDC. 

Figure~\ref{VLA-SMA-4} shows an image of the central region containing four of the 
five SMA dust cores. This region corresponds to the central 
 part of the massive clump MM1. The four panels show in: (a) the moment 0 map of the
 5 NH$_3$ (1,1) hyperfines, (b) the NH$_3$ (2,2) line, (c) the 4 NH$_3$ (1,1) 
satellites, and (d) the NH$_3$ (1,1) main component, in color-scale
 overlaid with black contours that correspond to the 1.3 mm dust continuum
 emission from the SMA.  
The NH$_3$ emission peaks do not overlap with the dust peaks, and remarkably, the
 NH$_3$ emission seems to avoid the dust cores (except in SMA5). This is more evident towards 
the SMA1 and SMA4 cores in panel (d). As can be seen in Figure~\ref{VLA-SMA-4}, the
 NH$_3$ emission 
 weakens toward the center of SMA1 and SMA4, independently of the transition used 
to make the moment map. This is likely produced for the combination of optical depth 
effects and depletion. The satellites could be self-absorved and become weak toward
 the densest parts in the cores. The NH$_3$ (2,2) may not be excited at the low temperatures 
 near de core's centers. In addition, at the low temperatures and high densities of the core's centers, 
 NH$_3$ could be frozen out onto dust grains. 
 
\subsubsection{Spectra}

\begin{figure*}
\begin{center}
\includegraphics[angle=0,scale=0.75]{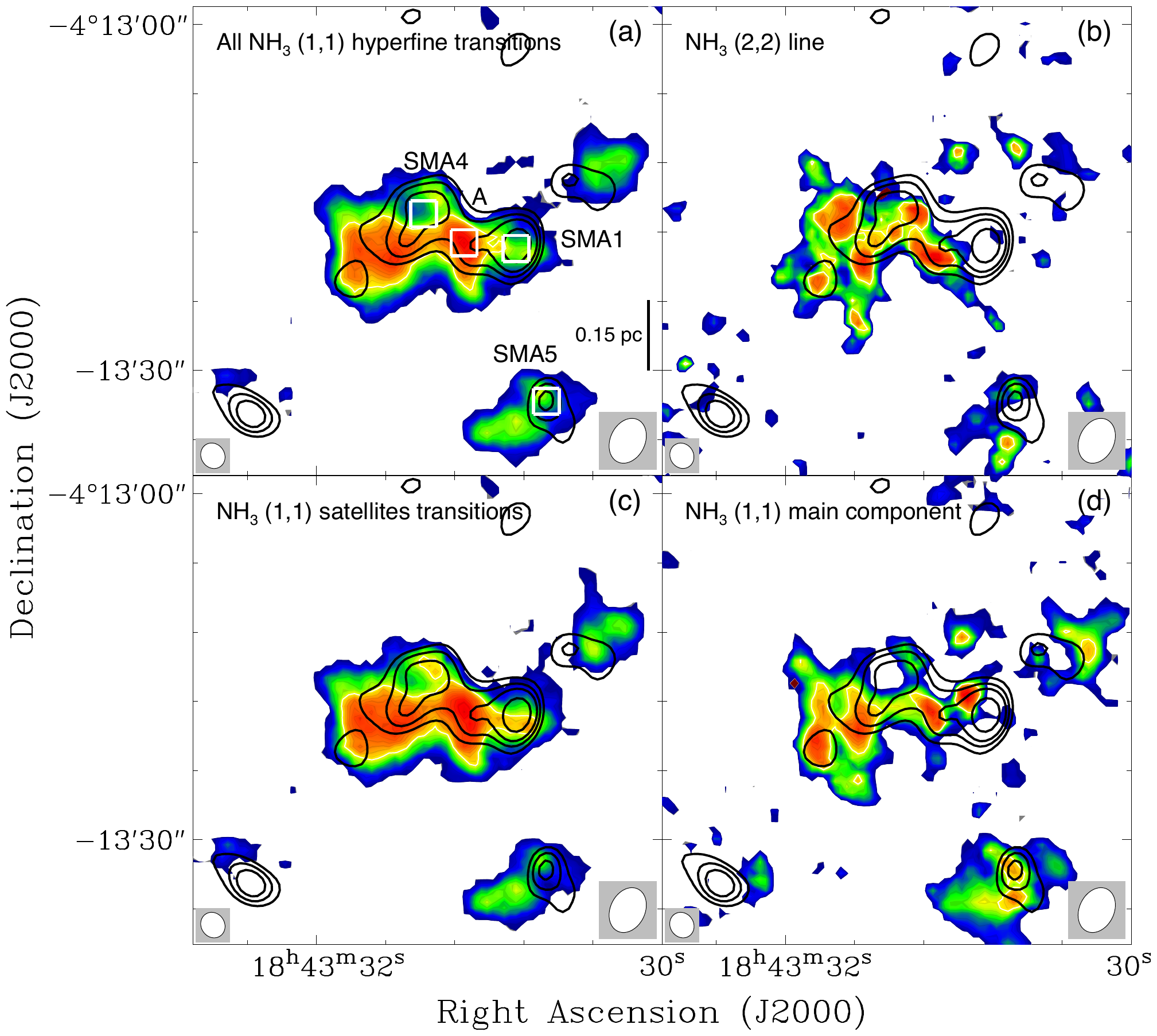}
\end{center}
\caption{NH$_3$ moment 0 map
 (integrated intensity) in
 color-scale  overlaid with 1.3 mm continuum emission from SMA in black
 contours toward the central region of the IRDC. Panel (a) corresponds to
 all five NH$_3$ (1,1) hyperfine lines. Panel 
(b) corresponds to the NH$_3$ (2,2) line. Panel (c) corresponds to the four
 NH$_3$ (1,1) satellites. Panel (d) corresponds to the NH$_3$ (1,1) main
 component. The moment 0 map includes the data above 2.5$\sigma$ per channel.
 All emission above 20\% of the peak emission is
 shown. Contour levels for the 1.3 continnum emission are 3, 4, 5, and
 7 $\times$ $\sigma$, with $\sigma$ equals to 0.75 mJ beam$^{-1}$.
 Angular resolutions of JVLA (2\farcs1) and SMA (3\farcs5) are shown in the
 bottom left and right, respectively. 
 White boxes show the positions of
 the spectra displayed in Figure~\ref{spectra}.}
\label{VLA-SMA-4}
\end{figure*}

\begin{figure}
\hspace{-0.8cm}
\includegraphics[angle=0,scale=0.55]{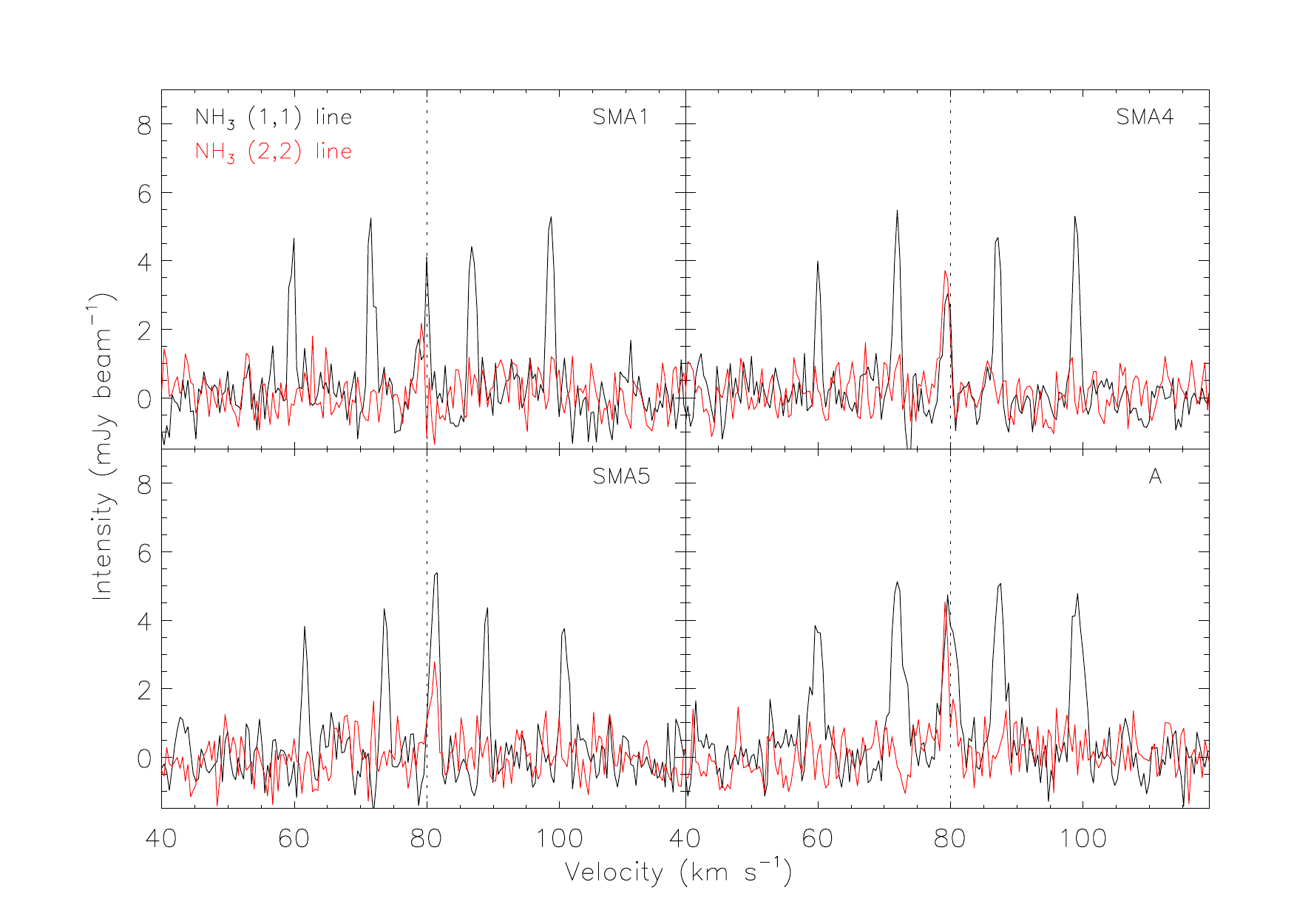}
\caption
{NH$_3$ (1,1) spectra (in black) toward the clump MM1 showing 
lines with high optical depths and self-absorption. NH$_3$ (2,2) spectra
 is showed in red for comparison. Position A is indicated
 in Figure~\ref{VLA-SMA-4}. The dotted line shows the $V_{LSR}$ of the
 IRDC (80 \kms). }
\label{spectra}
\end{figure}

Figure~\ref{spectra} shows the NH$_3$ (1,1) and (2,2) spectra toward selected positions
 in the clump MM1. At the position of the SMA5, the NH$_3$ (1,1) shows the 
 normal relative intensity between the main component
 and the four satellites, i.e., the main component brighter than the
 satellites. 
At the position of the SMA1 and SMA4 cores, the main component is weaker than the 
satellites, while at the intermediate position labeled as ``A'' in Figure~\ref{VLA-SMA-4}, 
the relative intensity among transitions is approximately unity.

The relative optical depths between the main NH$_3$ (1,1) component and the
 satellites is determined by quantum mechanics according to their statistical 
weights. However, the observed relative 
intensity can be modified by optical depth effects: in the optically thin limit, 
the relative intensity will equal the ratio of the statistical weights,
 whereas in the optically thick limit, the relative intensity will equal 1.
 Toward some 
positions in the clump MM1, the relative intensity of $\sim$1 likely indicates 
high optical depths. However, an additional explanation is required to
 explain why in some places the main hyperfine component is weaker than the
 satellites. This unique feature is likely produced by the self-absorption of the cold 
 gas in SMA1 and SMA4, as explained below.
 
 The critical density
  of NH$_3$ (1,1) is a few times 10$^4$ cm$^{-3}$. As 
 determined in Section~\ref{sma-mass}, the gas in the cores have
 densities of $\sim$10$^6$ cm$^{-3}$. At these high densities, the gas in
 the cores should be 
thermalized and LTE should hold. As will be discussed in
 Section~\ref{nh3-temp}, the 
temperature of the gas derived by using single-dish telescopes and
 interferometers points to a common gas temperature of $\sim$12 K. Unfortunately, a
 direct estimation of the temperature 
at the position of the SMA dust cores with the JVLA NH$_{3}$ data cannot be made because 
the main (1,1) component is less bright than its satellites. The emission from the main hyperfine 
is weaker than the satellites only at the position of the dust cores. This localized self-absorption 
rules out missing flux as a plausible explanation. Missing extended NH$_3$ emission would affect 
a much larger area and for no reason only specifically the dust cores. 

If a point-like warmer source 
is deeply embedded in the SMA cores, the surrounding, colder medium could absorb the warmer 
emission. As a result, one would observe a spectrum with a similar shape to the observed SMA spectrum 
at the core positions, except that emission lines should be generally brighter than the surrounding, 
colder medium (specially the satellites). In the clump MM1, we see that the satellite lines decrease 
in intensity toward the dust cores, which is opposite to the expected if there is an embedded warmer source. 

On the other hand, to explain the abnormal NH$_3$ profile, we suggest that the more diffuse 
(3$\times$10$^4$ cm$^{-3}$) gas/dust in the IRDC is at $\sim$12 K, whereas the material in the dense 
cores is colder ($<$12 K). Thus, from our line-of-sight, we see that the cold core gas absorbs the background 
warmer, more diffuse IRDC gas. In this picture, we assume that the surrounding medium is not symmetrically distributed 
around the cores and that there is an excess of background IRDC gas with respect to foreground IRDC gas. 
The peculiarity of the NH$_3$ spectra at the position of SMA1 and SMA4 would be the 
result of cold gas without internal heating sources, supporting the starless status of the cores.

\subsubsection{Line Widths\footnote{We refer to line width ($\Delta V$) to the full width half maximum (FWHM) 
of molecular line emission. The line width relates to the velocity dispersion ($\sigma$)  by $\Delta V$=2$\sqrt{2\ln 2}\,\sigma$$\approx$2.35$\,\sigma$.}}

Line widths are required to determine the turbulence of the gas, the virial mass, and the 
virial parameter. A multi-Gaussian function with fixed frequency separation
 between hyperfine transitions was fitted to an averaged spectra of 25 pixels 
(approximately the NH$_3$ beam size) centered on the SMA core positions. 
The line widths of the outer left hyperfine ($JKF_1=111\rightarrow 110$) 
associated with each SMA core are 
 displayed in Table~\ref{tbl-deltaV}. When line widths approach the
 thermal line width 
 ($\sim$0.2 \kms\ at 12 K), the magnetic hyperfine splitting can be resolved
\citep[e.g.,][]{Rydbeck77}, except for the outer left hyperfine. The two 
magnetic hyperfine transitions that form part of the outer left hyperfine are 
only separated by 0.14 \kms\ (11 kHz) and may be barely resolved in cold, 
low-mass star-forming regions. In IRDC G028.23-00.19, all other 
hyperfines show 20\% or larger 
 line widths than the one measured for the outer left hyperfine, indicating
 that the magnetic hyperfine splitting is becoming relevant at the line
 widths observed in this IRDC (although they are not resolved). 
 The observed line widths ($\Delta V_{\rm obs}$) of the outer left
 hyperfine displayed in Table~\ref{tbl-deltaV} have an average of 0.9 \kms,
 which is roughly 
 twice the channel width of the observations (0.4 \kms). The deconvolved
 line width ($\Delta V_{\rm dec}$) in \kms\ is determined following
 $\Delta V_{\rm dec}^2 = \Delta V_{\rm obs}^2 - 0.4^2$. Calculated values are
 displayed in Table~\ref{tbl-deltaV}. The deconvolved line width is
  10\% to 20\% lower than the observed line width and will be used
 for the determination of core's properties.

\begin{deluxetable*}{lccccc}
\tabletypesize{\normalsize}
\tablecaption{Measured and Derived Line Widths Associated to the SMA Cores\label{tbl-deltaV}}
\tablewidth{0pt}
\tablehead{
\colhead{Core} & \colhead{$\Delta V_{\rm obs}$} & \colhead{$\Delta V_{\rm dec}$} & \colhead{$\Delta V_{\rm int}$}& \colhead{$\Delta V_{\rm nt}$}& \colhead{Mach}\\
\colhead{Name} & \colhead{(\kms)}& \colhead{(\kms)}&\colhead{(\kms)} &\colhead{(\kms)}&\colhead{Number}\\
\colhead{(1)} & \colhead{(2)}& \colhead{(3)}&\colhead{(4)} &\colhead{(5)}&\colhead{(6)}\\
}
\startdata
SMA1 & 1.02  $\pm$ 0.15 & 0.91 $\pm$ 0.16 & $<$0.65 & 0.89 & 1.8\\
SMA2 & 0.68 $\pm$ 0.18 & 0.55 $\pm$ 0.22 & $<$0.39  & 0.52 &  1.1\\
SMA3 & 0.75 $\pm$ 0.25 & 0.63 $\pm$ 0.29 & $<$0.45  & 0.61 &  1.3\\
SMA4 & 0.94 $\pm$ 0.12 & 0.85 $\pm$ 0.13 & $<$0.60  & 0.83 &  1.7\\
SMA5 & 0.95 $\pm$ 0.17 & 0.86 $\pm$ 0.19 & $<$0.61  & 0.84 &  1.8\\
\enddata
\tablecomments{Columns (2), (3), and (4) correspond to line widths of the NH$_3$ outer left hyperfine. 
$\Delta V_{\rm obs}$ is the observed line width. $\Delta V_{\rm dec}$ is the deconvolved line width given by 
$\Delta V_{\rm dec}^2 = \Delta V_{\rm obs}^2 - 0.4^2$, where 0.4 \kms\ is the spectral resolution. $\Delta V_{\rm dec}$ 
will be used to calculate the physical parameters of the SMA cores. $\Delta V_{\rm int}$ 
is the intrinsic line width given by Equation~\ref{tau-correc}. 
Column (5) is the non-thermal component of the NH$_3$ determined from $\Delta V_{\rm dec}$, assumed to be the
 same as the non-thermal component of H$_2$. In column (6), the Mach number is calculated using
  $\Delta V_{\rm nt}$/$\Delta V_{\rm th}$, where $\Delta V_{\rm th}$ = 0.48 \kms\ corresponds to the H$_2$ thermal line width. } 
\end{deluxetable*}

The intrinsic line width ($\Delta V_{\rm int}$) of a line can be broadened
 by the line optical depth. As will be discussed in
 Section~\ref{nh3-temp}, the optical depth of the outer left hyperfine
 is at least $\sim$2.4 at the SMA core positions. Fitting a Gaussian
 profile to an optically thick or moderately thick line will result in an  
overestimation of the real velocity dispersion of the gas. To correct for
 this effect, the following expression can be used \cite[e.g.,][]{Beltran05}
\begin{equation}  
\frac{\Delta V_{\rm dec}}{\Delta V_{\rm int}}=\frac{1}{\sqrt{\ln 2}}~\sqrt{-\ln \left[-\frac{1}{\tau}\,\ln \left(\frac{1 + e^{-\tau}}{2}\right)\right]}~,
\label{tau-correc}
\end{equation}
where $\Delta V_{\rm int}$ profile is assumed to be Gaussian. Adopting $\tau$ = 2.4, the correction for 
optical depth produces intrinsic line widths $\sim$30\% smaller  
(see Table~\ref{tbl-deltaV}). We prefer to use $\Delta V_{\rm dec}$ over 
 $\Delta V_{\rm int}$ in the following calculations because the uncertainty in $\tau$ is large (which propagate 
 to $\Delta V_{\rm int}$, virial masses, and virial parameters).
  However, we stress that large optical depths will make $\Delta V_{\rm int}$ smaller, which 
 produces virial masses and virial parameters lower than the calculated with $\Delta V_{\rm dec}$.
 If $\Delta V_{\rm int}$ is used, our final conclusions do not change, they are rather reinforced. 

\section{Analysis}

\subsection{NH$_3$ Optical Depth and Rotational Temperature}
\label{nh3-temp}

The calculation of several physical parameters depends on the temperature of the medium. In this section, we 
 define the temperature that will be used for the calculation of physical parameters in the 
 following sections.
 
The optical depth of the NH$_3$ hyperfine lines can be derived by taking 
the ratio between the main and the satellite hyperfine components (assuming they have the same 
filling factor), following, for 
example, \cite{Sanhueza12}: 
\begin{equation}  
\frac{1-e^{-\tau_{_{(1,1,m)}}}}{1-e^{-\gamma \tau_{_{(1,1,m)}}}}=\frac{T_{\rm b_{(1,1,m)}}}{T_{\rm b_{(1,1,s)}}}~,
\label{tau-eqn-NH3}
\end{equation}
where $T_{\rm b_{(1,1,m)}}$ and $T_{\rm b_{(1,1,s)}}$ are the brightness temperatures of 
the main and satellites components, respectively. The factor $\gamma$ is the
  relative strength determined by the statistical weights.
 The value of $\gamma$ is 
1/3.6 for the two inner satellites and 1/4.5 for the two outer satellites
 \citep{Ho83,Rydbeck77}.

The rotational temperature, $T_R$, that characterizes the population
 distribution between the (1,1) and (2,2) states can be determined by 
\citep{Ho83}
\begin{equation}  
T_R = -\frac{41.5}{\ln \left[{0.282\left(\frac{\tau_{(2,2,m)}}{\tau_{(1,1,m)}}\right)}\right]}~\rm {K}~,
\label{TR-eqn}
\end{equation}
where $\tau_{(2,2,m)}$ can be determined from
\begin{equation}  
\frac{1-e^{-\tau_{_{(1,1,m)}}}}{1-e^{-\tau_{_{(2,2,m)}}}}=\frac{T_{\rm b_{(1,1,m)}}}{T_{\rm b_{(2,2,m)}}}~.
\label{tau-eqn-NH3}
\end{equation}

The determination of the optical depth and rotational temperature becomes  
unreliable when the the intensity ratio between $T_{\rm b_{(1,1,m)}}$ and
 $T_{\rm b_{(1,1,s)}}$ approaches unity and  impossible when the main NH$_3$
 component is weaker than the satellites. As discussed earlier, NH$_3$
 profiles with  these odd characteristics are repeatedly seen in 
IRDC G028.23-00.19, especially in the clump MM1. Thus, optical depths and
 rotational temperatures were determined only in a few positions and the mean
 values associated 
 with the SMA cores will be reported. These mean values were determined inside the contour defined by
  the 5$\sigma$ level in the dust continuum emission. The mean optical depth
   of the NH$_3$ (1,1) main component 
is 11, demonstrating that ammonia emission is optically thick. The mean 
optical depths of the NH$_3$ (1,1) satellites and NH$_3$ (2,2) main component 
are moderately optically thick with values of 2.4 and 1.2, respectively. 
All these optical depths values should be treated as lower limits for the 
SMA cores since, at the center of the cores, exact values cannot be 
determined due to the extremely high optical depths. 

The mean rotational temperature is 13 K. This value corresponds to an upper
 limit because at higher optical depths, the temperature is lower. 
 \cite{Dirienzo15} also estimate $\sim$13 K using lower angular resolution observations 
 of NH$_3$.  This
 temperature is within the uncertainties quoted by the other two methods used
 for temperature determination. Both the
 {\it Herschel} dust temperature and the temperature derived by the rotational
 technique using CH$_3$OH are 12 $\pm$ 2 K \citep{Sanhueza13}. This latter value, 12 $\pm$ 2 K,
  will be adopted for the rest of this work.

\subsection{Jeans Mass, Free Fall Time, and Dynamical Crossing Time of the MM1 Clump}
\label{JM-FF}

Here we calculate the Jeans mass to compare with the measured core masses in the MM1 clump. 
The comparison between the free fall time and the dynamical crossing time can give additional 
information on the dynamical state of the clump. 

If the fragmentation of a clump is governed by the Jeans instability, the initially homogeneous 
gas will fragment into smaller pieces defined by the Jeans length ($\lambda_J$) and the Jeans mass ($M_J$):
 \begin{equation}
\lambda_J = \sigma_{\rm th}\left(\frac{\pi}{G\rho}\right)~,
\label{jeanslength}
\end{equation}
 and
\begin{equation}
M_J = \frac{4\pi\rho}{3}\left(\frac{\lambda_J}{2}\right)^3=\frac{\pi^{5/2}}{6}\frac{\sigma_{\rm th}^3}{\sqrt{G^3\rho}}~,
\label{jeansmass}
\end{equation}
where $\rho$ is the mass density  and
 $\sigma_{\rm th}$ is the thermal velocity dispersion (or isothermal sound speed, $c_s$) given by
\begin{equation}
\sigma_{\rm th} = \left(\frac{k_{\rm B}T}{\mu m_{\rm H}}\right)^{1/2}~.
\label{sigma-th}
\end{equation}
The thermal velocity dispersion is mostly dominated by H$_2$ and He, and  it should be 
derived by using the mean molecular weight per free particle, $\mu=2.37$ \citep{Kauffmann08}. The thermal
 line width of the total gas 
 ($\Delta V_{\rm th}$ = 2$\sqrt{2\ln 2}\,\sigma_{\rm th}$) is 0.48 \kms. Assuming a mass density given by a 
 spherical clump of mass 1,500 \Msun\ and radius of 0.6 pc \citep{Sanhueza13}, the thermal Jeans length and mass of MM1 are 
  0.14 pc and 2.2 \Msun, respectively. 
   If we replace $\sigma_{\rm th}$ by the observed velocity 
dispersion ($\sigma_{\rm obs}$) in the MM1 clump, we can obtain the turbulent Jeans mass.  
Using the NH$_2$D (1-1) line width of 1.9 \kms\ ($\sigma_{\rm obs}$ = 0.81 \kms) observed on clump scales by 
 \cite{Sanhueza13}, we obtain a turbulent Jeans mass of the clump of 130 \Msun. 

The characteristic time for gravitational collapse (ignoring thermal pressure, turbulence, and magnetic fields),  
  known as the free fall time, is given by
\begin{equation}
t_{\rm ff} = \sqrt{\frac{3\pi}{32G\rho}}~,
\label{tff}
\end{equation}
and the dynamical crossing time, which depends on the radius of the clump ($R$) and its gas velocity 
dispersion ($\sigma_{\rm obs}$), is given by $t_{\rm dyn}$ = $R/\sigma_{\rm obs}$. 
The free fall time and the dynamical crossing time of MM1 are 2.0$\times$10$^5$ yr and  7.3$\times$10$^5$ yr, respectively.
A clump becomes gravitationally unstable if $t_{\rm ff}$ $<$ $t_{\rm dyn}$ \citep[e.g.,][]{Contreras17}. 
The $t_{\rm ff}$/$t_{\rm dyn}$ ratio is 0.3, indicating gravitational contraction.

\subsection{Non-thermal Component of the SMA Cores}
\label{TH}

The level of turbulence in the cores can be compare with the turbulence suggested by 
some high-mass star formation theories. 
Assuming that the NH$_3$ emission traces the velocity dispersion in the 
interior of the SMA cores, the non-thermal component ($\Delta V_{\rm nt}$)
 can be estimated using $\Delta V_{\rm dec}$ from the outer left hyperfine and the relation 
$\Delta V_{\rm dec}^2 = \Delta V_{\rm th}^2 + \Delta V_{\rm nt}^2$. The thermal line width
 ($\Delta V_{\rm th}$) for NH$_3$ is 0.18 \kms\ at 12 K. Because the non-thermal component
  is independent of the observed line from which is determined
 (except for molecular outflow/shock tracers), the determined $\Delta V_{\rm nt}$ from NH$_3$ 
represents the turbulent motions from the total gas in the cores (mostly H$_2$). $\Delta V_{\rm nt}$
 ranges from 0.52 to 0.89 \kms. The thermal line width of the total gas (for $\mu=2.37$) is 
0.48 \kms, leading to Mach numbers of ($\Delta V_{\rm nt}$/$\Delta V_{\rm th}$$\sim$1.1-1.8).
  If $\Delta V_{\rm int}$  is used instead of $\Delta V_{\rm dec}$, a lower non-thermal component 
  that is just 0.7-1.3 times the thermal line width is derived. $\Delta V_{\rm nt}$ and
 Mach numbers ($\Delta V_{\rm nt}$/$\Delta V_{\rm th}$) are given in Table~\ref{tbl-deltaV}. 

\subsection{SMA Cores' Properties using Dust Continuum Emission}
\label{sma-mass}

The core mass is determined to search for the existence of high-mass cores and compare 
with the thermal and turbulent Jeans masses. 
The mass of the cores was calculated by using the following expression:
\begin{equation}
M_{\rm core} = \mathbb{R}~\frac{F_\nu D^2}{\kappa_\nu B_\nu (T)}~,
\label{eqn-dust-mass}
\end{equation}
where $F_\nu$ is the measured integrated source flux, $\mathbb{R}$ is the
 gas-to-dust 
mass ratio, $D$ is the distance to the source, $\kappa_\nu$ is the dust
 opacity per gram of dust, and $B_\nu$ is the Planck function at the dust
 temperature $T$. A value of 0.9 cm$^2$ g$^{-1}$ is adopted for
 $\kappa_{1.3 mm}$, which corresponds to the opacity of dust grains with thin ice
 mantles at gas densities of 10$^6$ cm$^{-3}$ \citep{Ossenkopf94}.  
A gas-to-dust mass ratio of 100 was assumed in this work. The number density 
was calculated by assuming a spherical core and using the molecular mass per 
hydrogen molecule ($\mu_{\rm H_2}$) of 2.8 \citep{Kauffmann08}. Masses, number densities, 
and surface densities for all cores are listed in Table~\ref{tbl-SMA-cores-prop}.

 \subsection{Dynamical State of Embedded Cores}

The dynamical state of the cores is assessed by determining the virial mass and 
the virial parameter in order to compare with model predictions. 
The virial mass was evaluated according to the 
prescription of \cite{MacLaren88} (neglecting magnetic fields and external 
pressure):
\begin{equation}
M_{\rm vir} = 3\left(\frac{5-2n}{3-n}\right) \frac{R\sigma^2}{G}~,
\label{eqn-virial-mass2}
\end{equation}
  where $R$ is the radius of the core, $\sigma$ is the velocity dispersion
 along the line of sight, 
 $G$ is the gravitational constant, and $n$ is a constant whose exact value 
depends on the density profile, $\rho (r)$, as a function of the distance
 from the core center, $\rho (r) \propto r^{-n}$.  

Equation~\ref{eqn-virial-mass2} can be written in more useful units as:
\begin{equation}
M_{\rm vir} = k\left(\frac{R}{[pc]}\right)\left(\frac{\Delta
  V}{[\kms]}\right)^2\Msun~,
\label{eqn-virial-mass3}
\end{equation}
where $\Delta V$ is the line width and the value of $k$ depends on the density profile \citep{MacLaren88}. 
For a uniform density profile, $k$ is equal to 210. However, the uniform density profile is unlikely
 and represents an upper limit for the virial mass. In fact, \cite{Garay07} and
 \cite{Muller02} find, on average, a radial profile index of 1.8 in high-mass
 star-forming regions. The same average value for the radial profile index
 is also found on IRDC cores by \cite{Zhang09}. A density profile with
 $n=1.8$, resulting in $k=147$,  will be used in this work. 
The virial parameter ($\alpha$) was determined by taking the ratio between
 equation~\ref{eqn-dust-mass} and equation~\ref{eqn-virial-mass3}:
$\alpha = \frac{M_{\rm vir}}{M_{\rm core}}$. The calculated $M_{\rm vir}$ and 
$\alpha$ are listed in Table~\ref{tbl-SMA-cores-prop}.

\subsection{\normalsize Uncertainties in the Determination of Physical Parameters}
\label{sec:error}

There are several sources of uncertainty in the mass determination of
 star-forming regions. This section discusses the uncertainties associated
 to the parameters involved in the mass determination. 

\begin{deluxetable*}{lcccccccc}
\tabletypesize{\normalsize}
\tablecaption{Measured Properties of the SMA Cores\label{tbl-SMA-cores-prop}}
\tablewidth{0pt}
\tablehead{
\colhead{Core} & \colhead{$R$\tablenotemark{a}} & \colhead{$M_{\rm core}$} & \colhead{$M_{\rm core}$/$M_J$\tablenotemark{b}} & \colhead{n(H$_2$)} & \colhead{$\Sigma$} & \colhead{$\Delta V_{\rm dec}$} & \colhead{$M_{\rm vir}$\tablenotemark{c}} & \colhead{$\alpha$}\\
\colhead{Name} & \colhead{(pc)} & \colhead{(\Msun)} & \colhead{} & \colhead{($\times$10$^6$ cm$^{-3}$)}&\colhead{(g cm$^{-2}$)}&\colhead{(\kms)}&\colhead{(\Msun)}&\colhead{}\\
}
\startdata
SMA1 & 0.030 & 15  & 6.8 & 1.9  & 1.1   & 0.91 & 3.7 $\pm$ 1.4     & 0.25 $\pm$ 0.15 \\
SMA2 & 0.019 & 12  & ...   & 6.0  & 2.2   & 0.55 & 0.85 $\pm$ 0.70 & 0.07 $\pm$ 0.06  \\
SMA3 & 0.024 & 8.5 & 3.9 & 2.1  & 1.0   & 0.63 & 1.4 $\pm$ 1.3     & 0.2 $\pm$ 0.2   \\
SMA4 & 0.043 & 11  & 5.0 & 0.5  & 0.40 & 0.85 & 4.6 $\pm$ 1.5     & 0.41 $\pm$ 0.23 \\
SMA5 & 0.027 & 8.4 & 3.8 & 1.5  & 0.78 & 0.86 & 2.9 $\pm$ 1.3     & 0.35 $\pm$ 0.22 \\
\enddata
\tablecomments{Uncertainties for the radius ($R$), mass of the core
 ($M_{\rm core}$), volume density (n(H$_2$)), and surface density ($\Sigma$) are 10\%, 
49\%, 48\%, and 47\%, respectively (see discussion in Section~\ref{sec:error}). 
Uncertainty for $\Delta V_{\rm dec}$ is given in Table~\ref{tbl-deltaV}. 
$M_{\rm vir}$ and $\alpha$ correspond to the virial mass and virial parameter, 
respectively.}
\tablenotetext{a}{This radius ($R$) corresponds to half of the deconvolved size
 quoted in Table~\ref{tbl-SMA-cores}. For SMA4, the radius is half of the 
synthesized beam.}
\tablenotetext{b}{The Jeans mass of the clump MM1 is 2.2 \Msun\ (see Section~\ref{JM-FF}). SMA2 is not embedded in MM1.}
\tablenotetext{c}{Virial mass estimated by assuming a density distribution
 $\propto r^{-1.8}$.}
\end{deluxetable*}

The difficulty of characterizing interstellar dust makes $\kappa_\nu$  
the least known parameter for determining the core mass.  
In the literature, the models of \cite{Ossenkopf94} are broadly used and 
have been favored in multi-wavelength observations of star-forming regions 
\citep[e.g.,][]{Shirley11}. The value of $\kappa_\nu$ used in this work (0.9
 cm$^2$ g$^{-1}$) corresponds to the so-called OH5 grain which is covered 
 by a thin layer of ice mantle and coagulated at gas densities of 
10$^6$ cm$^{-3}$ \citep{Ossenkopf94}. The value of $\kappa_\nu$ ranges from 
0.7 to 1.05 if the values for thick layers and densities of 10$^5$ and 10$^7$
 cm$^{-3}$ are used instead. Assuming that the range of values is uniformly
 distributed between 0.7 and 1.05, the standard deviation can be determined
 by taking the size of the range (1.05-0.7 = 0.35) divided by $\sqrt{12}$. The value 
obtained is 0.1. Then, assuming 1-$\sigma$ uncertainty, $\kappa_\nu$ would be  
0.9 $\pm$ 0.1 (11\% uncertainty). \cite{Shirley11} constrain theoretical
 models of dust opacity at 450 and 850 $\mu$m. The OH5 model is one of three 
 supported by the observations. \cite{Shirley11} determine several values
 for $\kappa_{_{450 \mu\rm m}}$ and $\kappa_{_{850 \mu\rm m}}$. Assuming that the range of
 values they obtain is uniformly distributed between the extreme values,
 the standard deviation is 1.8 (28\% of $\kappa_{_{450 \mu\rm m}}=6.4$ from the
 OH5 model) and 0.34 (19\% of $\kappa_{_{850 \mu\rm m}}=1.8$ from the OH5 model). 
To be conservative, overall, a 1-$\sigma$ uncertainty of 28\% will be 
adopted for $\kappa_\nu=0.9$ at 1.3 mm. 

Another difficulty in determining the core mass is the conversion factor 
that relates the dust with the gas mass. The value of the canonical gas-to-mass
 ratio ($\mathbb{R}$) widely used is 100. Depending on the grain size,
 shape, and composition, determinations of the Galactic dust-to-gas mass
 ratio range between 70 and 150 \citep[e.g.,][]{Devereux90,Vuong03}. In this
 work, the canonical value of 100 has been adopted. Assuming that the range of
 $\mathbb{R}$ is uniformly distributed between 70 and 150, the standard
 deviation is 23. Thus, the 1-$\sigma$ uncertainty for the gas-to-mass
 ratio is 23 (23\% of $\mathbb{R}$=100). 

The dust temperature and measured continuum flux have an uncertainty of 17\%
 (see Section~\ref{nh3-temp}) and 15\% (see Section~\ref{sma-obs}),
 respectively. 
The major source of error in the kinematic distance method is the assumption of circular motions.
 Non-circular motions, e.g., cloud-cloud velocity dispersion (random motions), will lead to velocity perturbations
 of about 5 \kms. Using the rotation curve of
 \cite{Clemens85}, IRDC G028.23-00.19 is located at 5.1 kpc. Varying the
 velocity of G028.23-00.19 in $\pm$5 \kms, an uncertainty in the distance
 of 10\% is estimated. Another rotation curve places 
 the IRDC at a distance of 4.6 kpc \citep{Reid09}. In this work, the
 rotation curve of \cite{Clemens85} will be adopted; the distance derived
 by using the rotation curve of \cite{Reid09} agrees within the uncertainties. 

Due to their poor characterization, $\kappa_\nu$ and $\mathbb{R}$ add
 an ``intrinsic''  
uncertainty of 32\% to the mass determination of cores. Depending on 
how well the flux, distance, and temperature of the sources are determined, 
the uncertainty in the mass can be even higher than a factor 2. Because 
the SMA cores are observed with the same instrument, are at the same 
distance, and have the same temperature, all the SMA cores have 
the same mass uncertainty of 49\%. Due to the lower dependence 
on distance, the uncertainty for the volume density and surface density is 48\% and 
47\%, respectively. 

The uncertainty in the NH$_3$ line widths, and their effects on the virial
 mass 
and the virial parameter, will be different for each core. On average, line 
widths, virial masses, and virial parameters have 30\%, 60\%, and 75\%  
uncertainties. The uncertainties for each individual core are quoted in 
Table~\ref{tbl-SMA-cores-prop}. 

\section{Discussion}

\subsection{Implications for High-Mass Star Formation in IRDC G028.23-00.19}

\subsubsection{High-Mass Prestellar Cores}
      
         Recent works have focused on determining the mass of prestellar cores embedded in massive cluster-forming clumps 
         in order to test theoretical models \citep[e.g.,][]{Pillai11,Tan13,Zhang15}. However, the definition of a bonafide ``prestellar, high-mass core'' is rather vague. \cite{Longmore11} suggest that 
         in order to form an O type star through the direct collapse of a core, the core should have of the order of 100 \Msun. \cite{Tan14} suggest
          that prestellar, high-mass cores should have $\sim$100 Jeans masses. \cite{Krumholz07a} 
         simulate the formation a high-mass star of 9 \Msun\ from a turbulent, virialized core of 100 \Msun\ and 0.1 pc. It seems clear that 
         a prestellar, high-mass core should have several tens of solar masses. So far, no prestellar cores have been detected with such a large mass. 
         Indeed, follow-up observations of many suggested high-mass starless core candidates revealed molecular outflows or maser emission, irrefutable signs of star formation 
         \citep[e.g.,][]{Bontemps10,Duarte13,Tan13,Shipman14,Tan16,Feng16b}. 
         
         To be conservative, in this work we define a high-mass core as 
          a core with a mass larger than $\sim$30 \Msun. This definition is consistent with the star formation efficiency of 30\% derived by
           \cite{Alves07} in the Pipe dark cloud \citep[also tentatively determined in the Cygnus X complex by][]{Bontemps10}, assuming that the initial mass function 
           is a direct product of the core mass function as stated for the turbulent core accretion model, e.g., \cite{Tan14}. Interestingly, the prestellar core 
           candidate MM2 embedded in the active high-mass star-forming region G11.92-0.61 \citep{Cyganowski14,Cyganowski17} satisfies this condition
            and stands out as a good candidate to be a high-mass prestellar core.  We note that in our definition of high-mass core, 
           we have not considered that $\sim$80\% of high-mass stars are found in binary systems \citep{Kouwenhoven05,Chini12} and that the majority 
           of the systems contain pairs of similar mass.  
 
 The cores found in IRDC G028.23-00.19 have gas masses ranging from 8 to 15 \Msun. We therefore find no
 high-mass cores that currently have the mass reservoir to form a high-mass star, in disagreement with the core 
 accretion model.

\subsubsection{Fragmentation}

Dust and gas in  IRDC G028.23-00.19 MM1 seem to be cold and quiescent. 
\cite{Contreras17} find that massive clumps are more susceptible to 
gravitational instabilities and evolve faster than low mass clumps, based on the 
low $t_{ff}$/$t_{dyn}$ ratio. The low virial parameter and low $t_{ff}$/$t_{dyn}$ ratio indicate that 
MM1 has started gravitational contraction and is not a transient object. Indeed, 
at high angular resolution,  the first members of the future stellar cluster are 
revealed through dust continuum emission. At the 
sensitivity observed with SMA, no molecular outflows are detected and the cores
 embedded in this massive clump seem to be in the prestellar phase. 
Considering that the current observational evidence supports the idea that the IRDC will form 
high-mass stars, the lack of high-mass prestellar cores ($>$30 \Msun) have important 
implications in the formation of high-mass stars. 

\cite{Krumholz08} suggest that the heating produced by accreting low-mass stars 
in regions with surface densities $\geq$1 g cm$^{-2}$ can halt fragmentation by 
increasing the Jeans mass. Although surface densities of the order of 1 g cm$^{-2}$ are found 
at core scales in IRDC G028.23-00.19, only low temperatures are measured and there are no signs of active 
star formation. A similar conclusion was also reported in \cite{Zhang09} and \cite{Wang12} who studied a massive
 clump in IRDC G28.34+0.06 and found low gas temperatures of $\sim$14 K toward dense cores. 
In order to have a Jeans mass of 30 \Msun\ in IRDC G028.23-00.19 MM1, a temperature of 70 K 
would be needed. 
We note that heating from protostars seems unimportant even in G11.92-0.61, 
the high-mass star-forming region hosting a high-mass prestellar core candidate with a hot 
core nearby  \citep{Cyganowski14}. The measured temperature in the high-mass prestellar core candidate 
is 17-19 K. Magnetic fields have also been 
suggested as important to suppress fragmentation \citep{Commercon11,Myers13}. However, to date  
there are no measurements of the magnetic field in high-mass prestellar clump candidates, 
including IRDC G028.23-00.19. 

The Jeans mass in IRDC G028.23-00.19 MM1 is 2.2 \Msun. The observed SMA cores are 
 4--7 times more massive. The observational fact that core masses are larger 
than the Jeans mass is inconsistent with competitive accretion models, unless (i) we are 
not witnessing the initial fragmentation of the clump and the initial Jeans cores have had sufficient 
time to accrete and grow to reach their current masses or (ii) these cores could still fragment in smaller 
objects if higher angular resolution observations were available. We note that SMA observations are not 
sufficiently sensitive (5$\sigma$ = 4.5 \Msun) to detect Jeans cores, and we cannot exclude the possibility 
of a core population with lower masses separated by the Jeans length ($\lambda_J$). The Jeans length
 in MM1 is 0.14 pc and the separation among the SMA cores is larger than 2$\lambda_J$. 
 
 With the largest core mass of 15 \Msun, our 
observations disagree with the predictions from the turbulent core accretion model; high-mass 
prestellar cores are not found. 
The turbulent Jeans mass in IRDC G028.23-00.19 MM1 (130 \Msun) is much larger than the core 
masses (9 times larger than the most massive core). Therefore, contrary to other slightly more evolved 
IRDCs \citep[e.g.,][]{Wang11,Wang14,Lu15}, turbulence supported fragmentation does not seem to be the dominant 
process controlling the early stages of high-mass stars and cluster formation. Both the thermal Jeans mass and 
the turbulent Jeans mass may be too simple descriptions to explain the fragmentation of massive clumps. 
A larger sample will be key to confirm if this is a general trend at the very early stages of high-mass star 
formation evolution, or if IRDC G028.23-00.19 is a unique 
case where neither thermal pressure nor turbulent pressure dominate the fragmentation of a massive cluster-forming 
clump. 

\subsubsection{Turbulence}

The importance of turbulence can be further investigated by calculating the Mach number. 
As asserted by \cite{McKee03}, one of the most important premises of the
 turbulent core accretion model is that cores that will form high-mass stars
 are highly supersonically turbulent, leading to virial equilibrium
 ($\alpha$$\sim$1). 
NH$_3$ lines have narrow line widths in IRDC G028.23-00.19 ($\lesssim$1.0 \kms).  
 With Mach numbers ($\Delta V_{\rm nt}$/$\Delta V_{\rm th}$) of $\sim$1.1-1.8, the total gas is 
transonic and mildly supersonic. Although the gas may be slightly affected by turbulence,
 it is not highly supersonic (Mach numbers $>$5) as suggested by \cite{McKee03}, \cite{Krumholz07a}, and 
 \cite{Krumholz07b}.  
If optical depth is taken into consideration for the outer left NH$_3$ hyperfine, 
the $\Delta V_{\rm int}$ would produce even a lower non-thermal component that would be just 0.7-1.3
 times the thermal line width (subsonic-transonic).
 
 The simple analytic models developed by  \cite{Myers11,Myers14} describe how a protostar
gains mass from the collapse of a thermally supported core and from accretion of a turbulent clump.  
These models, based on statistical arguments, have a combination of ``core-fed'' and ``clump-fed''
 components, which represent isothermal collapse and reduced Bondi accretion. The duration of the
  accretion is more important than the initial core mass in settling the final mass of stars. \cite{Myers11,Myers14}
   suggest that the cores and the protostars that will become high-mass stars at the end of the cluster formation
    are born earlier than the low-mass counterparts from low-mass thermal cores. Stars become massive after accreting both thermal
  core gas and turbulent clump gas. Relatively low Mach numbers in the SMA cores in IRDC G028.23-00.19 may hint some similarity 
  to the work of \cite{Myers11,Myers14}. \cite{Myers14} suggests that 
  by the time that high-mass stars are identified, the core gas is turbulent because (i) its clump origin and (ii) the 
  star itself injects energy into the core (by heating, winds, and/or ionization). The SMA cores may 
  have already accreted a substantial amount of material from the clump, increasing the mass and turbulence in 
  agreement with \cite{Myers14}. However, we have no concrete evidence to support or refute that this scenario 
  is occurring in IRDC G028.23-00.19. 

\subsubsection{Dynamical State of the Cores}

The low turbulence level strongly affects the dynamics of the cores. Turbulent gas motions, and magnetic
 fields as well, can provide additional support against self-gravity. Considering only turbulence, all SMA cores consistently show $\alpha$$<$0.5,
  and are hence subvirial. These values suggest that in the absence of magnetic fields, the SMA cores are strongly 
  subvirial and simultaneously collapsing  along with the whole clump that has $\alpha$ = 0.3. This scenario of global 
  collapse dominated by subvirialized structures is consistent with some models of competitive accretion \citep[e.g.,][]{Wang10}, 
  and inconsistent with the turbulent accretion model \citep{McKee03,Tan14}. According to competitive accretion scenarios, 
  the low-intermediate mass cores in IRDC G028.23-00.19 could grow in mass by accreting gas from a
reservoir of material in the molecular cloud that may not be bound to any core. SMA1 and SMA4 would be the primary candidates 
to form high-mass stars in the future due to their position inside the clump, apparently near the center of the gravitational potential.  
On the other hand, the assumptions initially made in the turbulent core accretion model may not be applicable and may need to be
 reconsidered to represent better the observations. 

However, so far in the discussion, magnetic fields have been ignored and 
they can add additional support against collapse. Recently, \cite{Zhang14} obtained dust polarization information toward a
  sample of 14 massive star-forming regions. They found that magnetic fields in dense cores tend to follow the field orientation
   in their parental clumps. Therefore, they suggest that the magnetic field plays an important role in the fragmentation of clumps and the formation
    of dense cores. If magnetic fields are 
included in the virial equation, the following expression holds 
\begin{equation}  
M_{B, \rm vir} = 3\,\frac{R}{G}\left(\frac{5-2n}{3-n}\right)\left(\sigma^2 + \frac{1}{6}\sigma_A^2\right)~,
\label{virial_B}
\end{equation}
where $\sigma_A$ is the Alfven velocity, and $n$ depends on the density profile (as in Equation~\ref{eqn-virial-mass2}). 
The Alfven velocity can be determined from
\begin{equation} 
\sigma_A = \frac{B}{\sqrt{4\pi \rho}}~,
 \label{alfven_B}
\end{equation}
where $B$ is the magnitude of the magnetic field and $\rho$ is the mass density of each core, $\rho$ = $\mu_{H_2}\, m_{\rm H}\, {\rm n(H_2)}$.

To maintain virial equilibrium including magnetic fields
 ($M_{B, \rm vir}/M_{\rm core}=1$), field strengths of 1.7, 1.6,
 0.56, and 1.2 mG are needed towards the dust cores embedded in MM1 
 (SMA1, SMA3, SMA4, and SMA5, respectively). 
 If a uniform density is assumed, $n=0$ instead of $n=1.8$, the magnetic
 field magnitudes are $\sim$25\% lower. On 
 average, magnetic fields of $\sim$1-2 mG would be needed to maintain the SMA 
cores in virial equilibrium. Magnetic fields of these strengths are apparently 
 consistent with observations of more evolved high-mass cores. \cite{Crutcher10} suggest that at 
densities of $\sim$10$^6$ cm$^{-3}$ the most probable maximum strength for 
the magnetic field is $\sim$1 mG. \cite{Girart13} indeed measure the 
magnitude of the magnetic field toward the high-mass star-forming core DR
 21(OH) and determine a value of 2.1 mG at a density of 10$^{7}$ cm$^{-3}$. 
Although it seems possible to obtain magnetic field magnitudes of $\sim$1 mG
 in high-mass star-forming cores \citep[e.g.,][]{Qiu14,Li15}, so far, all estimation of field strengths are 
in cores with current evidence of star formation. No measurements have been 
made in prestellar sources. Therefore, although the non-magnetized version
 of the turbulent core accretion model is not consistent with the observed
 properties of the SMA cores in IRDC G028.23-00.19, which are candidates to form high-mass stars,
 the magnetized picture still needs to be tested. 

If observations indeed prove that the magnetized picture is feasible, it would imply that: 
(i) star formation efficiencies much larger than 30\% would be needed to form high-mass stars 
in IRDC G028.23-00.19 \citep[30\% is the current value most favor by observations, but higher values
 have been suggested from simulations, e.g.,][]{Matzner00} or (ii) 
IRDC G028.23-00.19 may never form high-mass stars, which would create a new puzzle 
and it would be necessary to understand why.

The largest possible mass that can be supported by a magnetic field is given by \citep{Bertoldi92}
\begin{equation}  
M_B = 16.2 \left(\frac{R_e}{Z_e}\right)^2  \left(\frac{n(\rm H_2)}{10^6\,\, {\rm cm^{-3}}}\right)^{-2}  \left(\frac{B}{mG}\right)^3 \Msun~,
\label{M_B}
\end{equation}
where $2Z_e$ is the length of the symmetry axis and $R_e$ is the radius normal to the axis of an ellipsoidal core. 
Assuming a spherical core ($Z_e$=$R_e$) and $M_B$ equal to the mass of SMA1 (15 \Msun), we find the minimum
 magnetic field (1.5 mG) that would suppress fragmentation. This magnetic field strength is practically the same 
 that the field strength necessary to maintain the SMA1 core in virial equilibrium (as is the case also for the other cores embedded in the 
 clump MM1). If lower magnetic field strengths are eventually measured, the cores may be prone to fragment. 

 \section{Conclusions}
 
 We have imaged the IRDC G028.23-00.19 with the SMA ($\sim$3\farcs5 at 224 GHz) and
  JVLA ($\sim$2\farcs1 at 23 GHz). This IRDC hosts a massive (1,500 \Msun), cold (12 K), and 
  IR dark (at {\it Spitzer} 3.6, 4.5, 8.0, and 24 $\mu$m and at {\it Herschel} 70 $\mu$m) clump, which is 
  one of the most massive, quiescent clumps known (MM1). After examining the dust continuum 
  and the spectral line emission, we draw the following conclusions: 
  
  1. Using the SMA dust continuum emission, 5 dense cores are detected: 4 of them are embedded in 
  MM1 (SMA1, SMA3, SMA4, and SMA5) and one is located in the northern part of the IRDC (SMA2).
  There are no cores with a mass larger than 15 \Msun. The lack of high-mass prestellar cores is in 
  disagreement with the turbulent core accretion model. In order to form a high-mass star, the SMA1 core 
  needs, at least, to double its mass at the same time the central star accretes material. 
The idea that high-mass stars can form without passing through a high-mass stage in the prestellar phase
 is consistent with competitive accretion scenarios \citep[e.g.,][]{Bonnell04, Wang10,Myers11,Myers14}. 
  
  2. The Jeans mass (2.2 \Msun) is 4 to 7 times smaller than the core's masses. This is in disagreement with the 
  prediction of competitive accretions models where the clumps fragment in objects with masses similar 
  to the Jeans mass,  unless the SMA cores have had 
  sufficient time to accrete and significantly increase their mass.
  
  3. Neither CO wing emission or SiO emission was detected indicating molecular outflows. 
  Thus, confirming that, at the sensitivity of these observations, the SMA cores are starless. To our 
  knowledge, IRDC G028.23-00.19 MM1 is the most massive, cold clump that after interferometric observations 
  maintains the status of ``prestellar candidate."
  
  4. At core scales, the NH$_3$ line widths have some contribution from turbulence, with Mach 
  numbers ranging from 1.1 to 1.8. The gas in the SMA cores is not highly supersonic as the turbulent 
  core accretion suggests. 
  
  5. By comparing the thermal and turbulent Jeans masses with the SMA core's masses, we find that the 
  global fragmentation of the clump MM1 is dominated by neither thermal nor turbulent pressure. 
  
  6. Unless magnetic fields strengths are about 1-2 mG, the cores are strongly subvirialized ($\alpha$ $<$ 0.5). 
  The SMA cores are significantly below equilibrium and likely under fast collapse, which is consistent with cores 
  that can grow in mass. 
  
 7. We finally conclude that in IRDC G028.23-00.19 we are witnessing the initial fragmentation of a massive, prestellar clump that 
 will form high-mass stars. 
 Whether the properties observed in IRDC G028.23-00.19 are unique or typical of the very early stages of high-mass star formation 
 needs to be confirmed with a larger well-defined sample.  
  
  \acknowledgements
  
  P.S. gratefully acknowledge Jonathan B. Foster, Satoshi Ohashi, and Fumitaka Nakamura for helpful 
  discussions. P.S. thanks the comments from the anonymous referee. 
A.E.G. thanks FONDECYT N$^{\rm o}$ 3150570. K.W. is supported by grant WA3628-1/1 of the German
 Research Foundation (DFG) through the priority program 1573 (``Physics of the Interstellar Medium''). 
 Data analysis was in part carried out on the open use data analysis computer system 
at the Astronomy Data Center, ADC, of the National Astronomical Observatory of Japan.

%

\vspace{5mm}
\facilities{SMA, JVLA}


\software{IDL, MIR, CASA
          }

 \appendix
 \section{Derivation of the maximum stellar mass using the IMF}
\label{app}

Adopting the IMF from \cite{Kroupa01}, we have $\xi(m) \propto$ $m^{-1.3}$
 for 0.08 \Msun\ $\leq$ $m$ $<$ 0.5 \Msun\ and $\xi(m) \propto$ $m^{-2.3}$ for 
 $m$ $\geq$ 0.5 \Msun, where $m$ corresponds to the star's mass and 
 $\xi(m) dm$ is the number of stars in the mass interval $m$ to $m + dm$. 
Assuming a range of stellar masses between 0.08 and 150 \Msun, we can impose 
the total number of stars with $m \geq m_{\rm max}$ to be unity (in order to assure the 
formation of one high-mass star, the lowest value for $m_{\rm max}$ should be 8 \Msun), 
\begin{equation}
1 = \int_{m_{\rm max}}^{150} \xi(m)\, dm~.
\label{eqn-IMF-unity}
\end{equation}

The total mass in a stellar cluster, $M_{\rm  cluster}$, is given by
\begin{equation}
M_{\rm  cluster} = \int_{0.08}^{150} \xi(m)\, m\, dm~.
\label{eqn-IMF-cluster-mass}
\end{equation}

Combining equation~\ref{eqn-IMF-unity} and~\ref{eqn-IMF-cluster-mass}
\begin{equation}
M_{\rm  cluster} = \frac{ \int_{0.08}^{150} \xi(m)\, m\, dm}{\int_{m_{\rm max}}^{150} \xi(m)\, dm}~,
\label{eqn-IMF-com}
\end{equation}
 and assuming a star formation efficiency, $\epsilon_{\rm sfe}$, of 30\% ($M_{\rm cluster}=0.3\times M_{\rm clump}$), we can 
relate $m_{\rm max}$ with the clump mass as 
\begin{equation}
m_{\rm max} = \left(\frac{0.3}{\epsilon_{\rm sfe}}\frac{17.3}{M_{\rm clump}} + 1.5\times 10^{-3}\right)^{-0.77}~.
\label{eqn-IMF-m-max-ap}
\end{equation}

To estimate the necessary mass in a clump to form a high-mass star, we can use the 
following relationship 
\begin{equation}
M_{\rm clump} = \frac{0.3}{\epsilon_{\rm sfe}}\,\, \frac{17.3}{(m_{\rm max}^{-1.3} - 1.5\times10^{-3})}~.
\label{eqn-IMF-m-clump}
\end{equation}

and making $m_{\rm max}$ = 8 \Msun\ we obtain 260 \Msun.

\end{document}